\documentclass[11pt,a4paper]{article}
\usepackage{amsmath}
\usepackage{authblk}
\usepackage{graphicx}
\usepackage{color}
\usepackage{latexsym}
\usepackage{geometry}
\newcommand{\hrd}[1] {{#1}}

\begin{document}

\title{Active quenching of superconducting nanowire single photon detectors}

\author[1]{Prasana Ravindran}
\author[2]{Risheng Cheng}
\author[2]{Hong Tang}
\author[1,3]{Joseph C. Bardin}
\affil[1]{University of Massachusetts Amherst}
\affil[2]{Yale University}
\affil[3]{Google, LLC}
\date{\vspace{-5ex}}
\maketitle



\begin{abstract}
Superconducting nanowire single photon detectors are typically biased using a constant current source and shunted in a conductance which is over an order of magnitude larger than the peak normal domain conductance of the detector. While this design choice is required to ensure quenching of the normal domain, the use of a small load resistor limits the pulse amplitude, rising-edge slew rate, and recovery time of the detector. Here, we explore the possibility of actively quenching the normal domain, thereby removing the need to shunt the detector in a small resistance. We first consider the theoretical performance of an actively quenched superconducting nanowire single photon detector and, in comparison to a passively quenched device, we predict roughly an order of magnitude improvement in the slew rate and peak voltage achieved in this configuration. The experimental performance of actively and passively quenched superconducting nanowire single photon detectors are then compared.  It is shown that, in comparison to a passively quenched device, the actively quenched detectors simultaneously exhibited improved count rates, dark count rates, and timing jitter.
\end{abstract} 

\section{Introduction}
Superconducting nanowire single photon detectors (SNSPDs) have gained widespread use due to their near saturated quantum efficiency\cite{MarsiliF,kahl2015waveguide,verma2015high}, high count rates\cite{rosenberg2013high,Robinsonoptics}, exceptional timing jitter\cite{14p8Jitter,zadeh2018single,caloz2018high,JPL}, and low dark count rates\cite{schuck2013waveguide,wollman2017uv}. The performance of an SNSPD is impacted by its bias and readout circuitry, and several research groups have worked on the modeling and optimization of this interface. An equivalent circuit describing the electrical interaction between an SNSPD and a passive termination was first described by Kerman and it was shown that the recovery time of a passively quenched SNSPD is limited by an electrical time constant, $\tau_\text{e}=L_\text{K}/R_\text{L}$, where $L_\text{K}$ is the kinetic inductance of the detector and $R_\text{L}$ is the load resistance\cite{kerman2006kinetic}.  Shortly thereafter, finite-difference time-domain simulations of the electro-thermal response of a passively terminated SNSPD were leveraged to study the extent to which the recovery time of a SNSPD with fixed kinetic inductance can be reduced by increasing $R_\text{L}$. It was found that only limited improvement was possible due to latching, which describes the condition in which a SNSPD becomes stuck in the resistive state~\cite{Latching}. This result was verified experimentally and it was shown that latching occurs when electro-thermal feedback is able to stabilize the normal domain, which happens when the ratio of the electrical to thermal time constant is too small\cite{KERMANFB}. 

While short detectors are desired to minimize the recovery time, a large area detector is typically employed to maximize system detection efficiency. As such, various approaches have been developed to minimize the reset time while preserving other performance characteristics. These techniques include biasing the detector through a resistor while employing a dc-coupled amplifier for readout~\cite{kerman2013readout}, leveraging a snubber element (consisting of a short circuited transmission line) to limit the charging and discharging of an ac-coupling capacitor\cite{snubber,improvedreadout}, and gating the detector using a microwave bias\cite{akhlaghi2012gated}. However, in each of these cases, the output voltage is limited to the product of the load resistance (typically 50$\,\Omega$) and the bias current ($I_\text{B}$). As the peak resistance that a normal domain within a passively quenched detector typically reaches during a detection event is on the order of twenty times that of the load resistance, a significant gain in pulse amplitude could be obtained if the detector were configured to operate with a high-impedance load. Increasing the pulse amplitude has the potential to reduce timing jitter due to the associated improvement in signal to noise ratio. 

In this paper, we explore the use of an active quenching architecture for the bias and readout of SNSPDs. We begin with the details of operation and the expected performance. We then describe a prototype active quenching system and present experimental results that demonstrate the advantages of this approach over conventional readout techniques. These results confirm that the count rate, dark count rate, and timing jitter of an SNSPD can be simultaneously enhanced through the use of active quenching. 

\section{Proposed approach}
\begin{figure}[bt!]
\centering
\includegraphics[width=0.75\columnwidth]{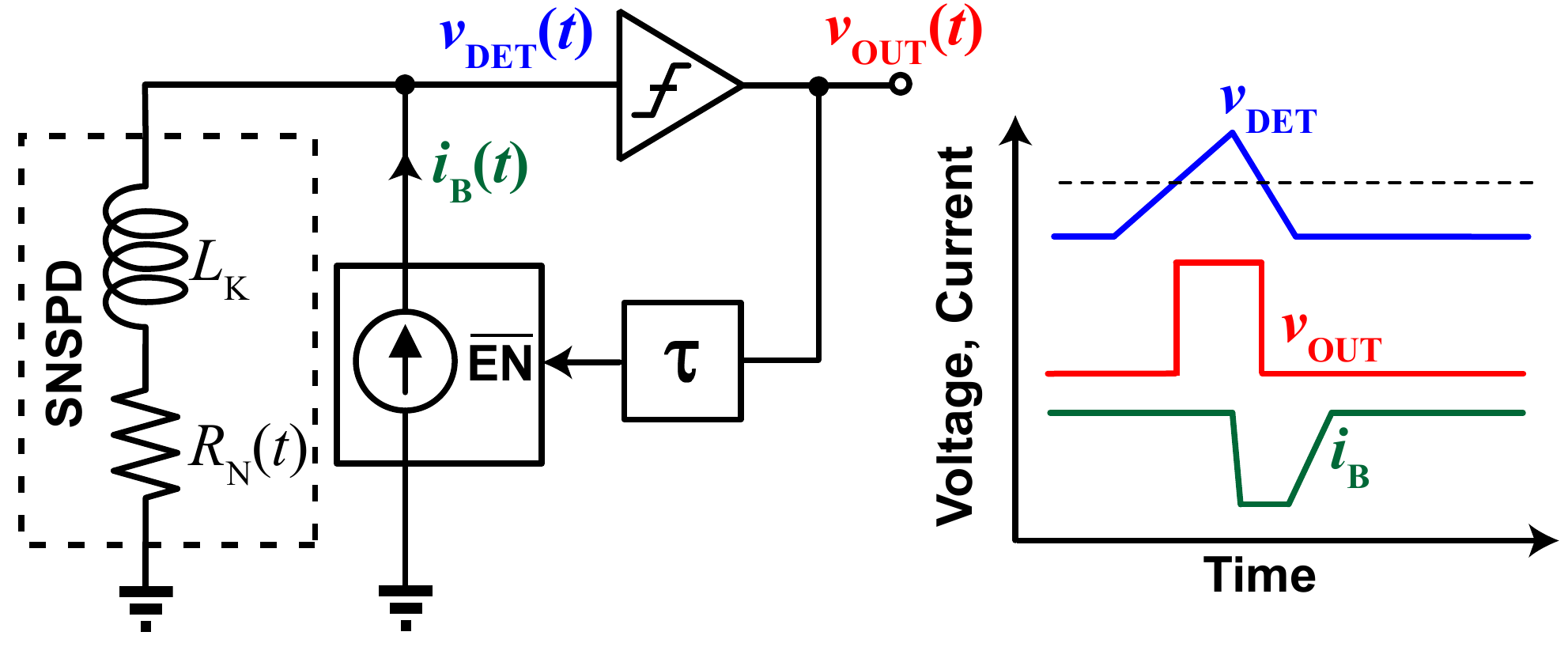}
\caption{Conceptual block diagram of active quenching architecture and key waveforms describing its operation.}
\label{concept}
\end{figure}

A block diagram of an actively quenched SNSPD appears in Fig.~\ref{concept}. The detector is biased using a constant current source, which is disabled whenever the voltage across the device exceeds the threshold of a comparator. Thus, once a normal domain forms due to a detection event or a dark count, it will grow until the aggregate resistance ($R_\text{N}$) is large enough that the comparator is triggered. At this point, the bias current will be disabled, allowing the normal region of the device to transition back to the superconducting state. After a delay, the bias current is  again enabled, returning the SNSPD to the photosensitive state. 

In principle, this approach has two significant advantages over standard SNSPD readout techniques. First, both the amplitude of the voltage across the detector 
as well as the slew rate associated with the rising edge of this voltage will be greatly enhanced since the SNSPD is no longer shunted by a small resistance. As the jitter added by the readout circuit is inversely proportional to slew rate, it is expected that the system timing jitter  will be enhanced.  Secondly, by actively quenching the normal domain, we decouple the reset and rebias operations, allowing for reduced dead times, corresponding to higher count rates. 
\section{Theoretical performance}
\begin{figure}[bt!]\centering
\includegraphics[width=\columnwidth]{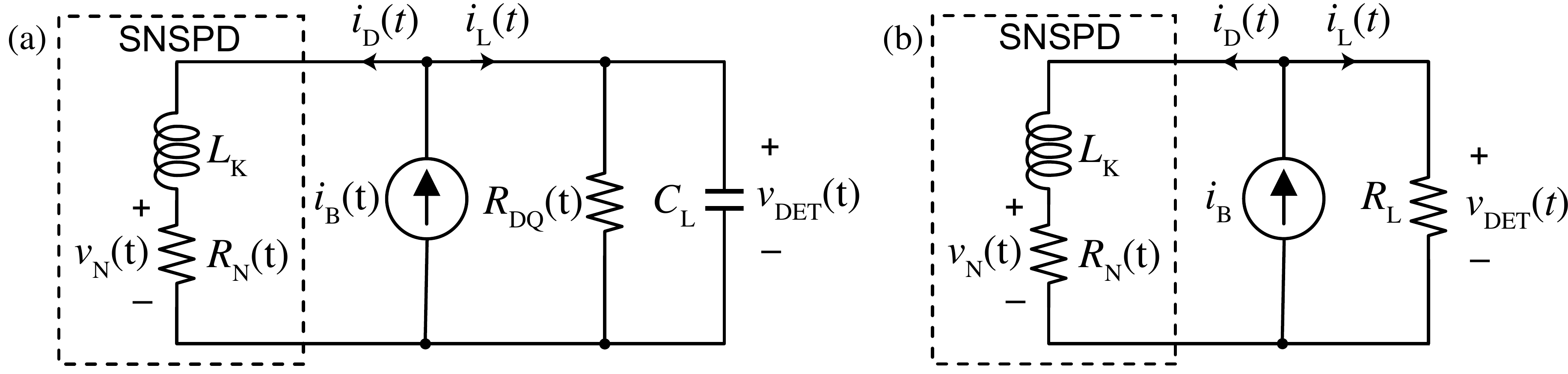}
\caption{Equivalent circuit of an (a) actively quenched and (b) passive quenched SNSPD.}
\label{SCH}
\end{figure}

The rising-edge performance of the proposed architecture can be quantified by considering the dynamics of normal domain growth within a capacitively loaded SNSPD
\footnote{For the purpose of the analysis described in this work, we consider an SNSPD as a lumped device and neglect the non-linearity of the kinetic inductance~\cite{clem2012kinetic}. While this is valid for short SNSPD, a distributed model should technically be used when studying devices with large kinetic inductance\cite{santavicca2016microwave}. However, based on measurement results, we believe our simplified models are sufficient to predict performance, even when considering larger devices.}, as shown in Fig.~\ref{SCH}(a). For this analysis, we assume that the bias current is kept constant ($I_\text{B}\left(t\right)=I_\text{B0}$) and that $R_\text{DQ}$ is infinite. Our goal is to estimate the improvement in slew rate ($\mathrm{d}v_\text{DET}/\mathrm{d}t$) and pulse amplitude that is achieved through the active quenching circuit, relative to the case in which the detector is biased from a constant current source and shunted in a load resistance ($R_\text{L}$), as shown in Fig.~\ref{SCH}(b).

To begin, we note that, when subjected to a current greater than that required to sustain a self-heating normal domain ($I_\text{D}>I_\text{SS}$)
, the normal domain can be described by a non-linear capacitance of the form\cite{SUSTKB,KERMANFB} $C_\text{eff}\left\{I_\text{D}\right\}\approx{}wd\sqrt{\psi{}I_\text{D}^2/I_\text{SW}^2-1}/\left[\left(2\rho_\text{n}v_\text{0}\right)\left(\psi{}I_\text{D}^2/I_\text{SW}^2-2\right)\right]$, where $I_\text{SW}\le{I_\text{C}}$ is the switching current, $v_\text{0}\equiv\sqrt{\left(h_\text{c}\kappa\right)/\left(c^2d\right)}$ is the characteristic normal domain velocity, $\psi\equiv\rho_\text{n}I_\text{SW}^2/\left[\left(h_cw^2d\right)\left(T_\text{C}-T_\text{SUB}\right)\right]$ is the Stekly parameter~{\cite{stekly}}, $T_\text{SUB}$ is the bath temperature, $h_\text{c}$ is the thermal boundry conductance between the superconducting film and the substrate, $w$ is the nanowire width,  and $d$, $\rho_\text{n}$, $\kappa$, $c$, and $T_\text{C}$ are the thickness, normal resistivity, thermal conductivity, specific heat per unit volume, and critical temperature of the superconducting film, respectively. 

Next,  we assume that a normal domain of infinitesimal length bridges the nanowire at $t=0$. All of the bias current will still be flowing through the nanowire ($I_\text{D}\left(0\right)=I_\text{B0}$) and the voltage across the normal domain will begin to slew as $\mathrm{d}v_\text{n}/\mathrm{d}t=I_\text{B0}/C_\text{eff}\left\{I_\text{B0}\right\}$. As this voltage grows, the current through the nanowire will decrease as part of the bias current is diverted to charge the load capacitance. Eventually, at time $t_\text{D}$, the currents charging the normal domain and the load capacitance will equalize such that the voltages across the normal domain and the load capacitance slew at the same rate, i.e., when 
$I_\text{SLEW}C_\text{L}=\left(I_\text{B0}-I_\text{SLEW}\right)C_\text{eff}\left\{I_\text{SLEW}\right\},$ 
where $I_\text{SLEW}$ is the steady-state current flowing through the nanowire. So, between $t=0$ and $t=t_\text{D}$, the kinetic inductance will have discharged by  $\Delta{E_\text{L}}=\left(I_\text{B0}-I_\text{SLEW}\right)^2L_\text{K}/2$. As this discharge aids normal domain growth, the slew rate will be maximum during this time interval. Thus, we consider the time interval during which the inductor discharges as that of the highest performance and constrain the circuit operation such that a reset is triggered prior to $t_\text{D}$. 

By approximating $C_\text{eff}$ as a linear function of time, it can be shown that the relative improvements in slew rate and signal amplitude achieved by active quenching are approximated as \hrd{(a derivation is given in Appendix A)}
\begin{equation}
\frac{SR_\text{AQ}}{SR_\text{PQ}}\approx\frac{2\sqrt{2L_\text{K}C_\text{eff}\left\{I_\text{B0}\right\}}}{R_\text{L}\left(\left<C_\text{eff}\right>+C_\text{L}\right)}
\label{SRcomp}
\end{equation}
and
\begin{equation}
\frac{v_\text{PK,AQ}}{v_\text{PK,PQ}}\approx\sqrt{2.2\frac{I_\text{SLEW}}{I_\text{B0}}\frac{L_\text{K}/R_\text{L}}{R_\text{L}\left(\left<C_\text{eff}\right>+C_\text{L}\right)}},
\label{voltcomp}
\end{equation}
where $\left<C_\text{eff}\right>=\left(C_\text{eff}\left\{I_\text{B0}\right\}+C_\text{eff}\left\{I_\text{SLEW}\right\}\right)/2$. For  parameter values consistent with a NbN detector on a sapphire substrate
 that is biased at 0.93\,$I_\text{SW}$, Equations~(\ref{SRcomp}) and (\ref{voltcomp}) evaluate to unity for kinetic inductances of approximately 200 and 100\,pH, respectively\footnote{Here we have assumed a load capacitance of 50\,fF, corresponding to $I_\text{SLEW}\approx{}I_\text{SW}/2$.}. For larger detectors, the improvement scales proportional to the square root of the kinetic inductance. 


It is also important to understand the requirements that must be met for the  detector to recover and be re-biased. First, we consider the operation of resetting the detector to the superconducting state. For the detector to recover, $I_\text{D}$ must drop below $I_\text{SS}$ so that phonon cooling can overcome Joule heating. 
However, since recovery occurs at a non-zero value of $I_\text{D}$, the kinetic inductance is still energized ($E_\text{L}\approx{I_\text{SS}^2L_\text{k}/2}$). Referring to Fig.~\ref{SCH}(a), when the detector is in the superconducting state, the circuit behaves as a parallel LC resonator with resonant frequency $\omega_0=1/\sqrt{L_\text{K}C_\text{L}}$ and quality factor $Q=R_\text{DQ}\sqrt{C_\text{L}/L_\text{K}}$. To prevent ringing of the voltage at the output of the detector, it is desirable for $R_\text{DQ}$ to be smaller than $\sqrt{L_\text{K}/C_\text{L}}/2$ during the reset operation. \hrd{Similarly, when the detector is rebiased, it is essential that the circuit is sufficiently damped to prevent ringing, which could cause the instantaneous bias current through the detector to exceed $I_\text{SW}$, leading to oscillation. However, in this case,} it is feasible to employ a somewhat larger value of $R_\text{DQ}$ if the slew rate of the bias current waveform is limited (so as to control the energy at frequencies in the vicinity of $\omega_\text{0}$). \hrd{Thus,} one may choose to tolerate some ringing after the quench operation so that a single value of $R_\text{DQ}$ may be used while the device is quenched and rebiased.


From the discussion above, it is clear that the performance of an actively quenched SNSPD is expected to be optimum when the capacitance loading the detector is minimized. Specifically, it is desirable that the total capacitance seen at the output node of the detector is at most of the same order of magnitude as that of $\left<C_\text{eff}\right>$. For a typical NbN detector on a sapphire substrate, $\left<C_\text{eff}\right>$ is on the order of 35\,fF. As the bondpads of a typical CMOS integrated circuit present a capacitance on the order of 30\,fF, it is essential that other capacitances presented by the readout and interface circuitry be minimized. It is therefore important to tightly integrate the SNSPD with the active quenching circuit so as to minimize parasitic capacitances.

\section{Results}

\begin{figure}[bt!]\centering
\includegraphics[width=\columnwidth]{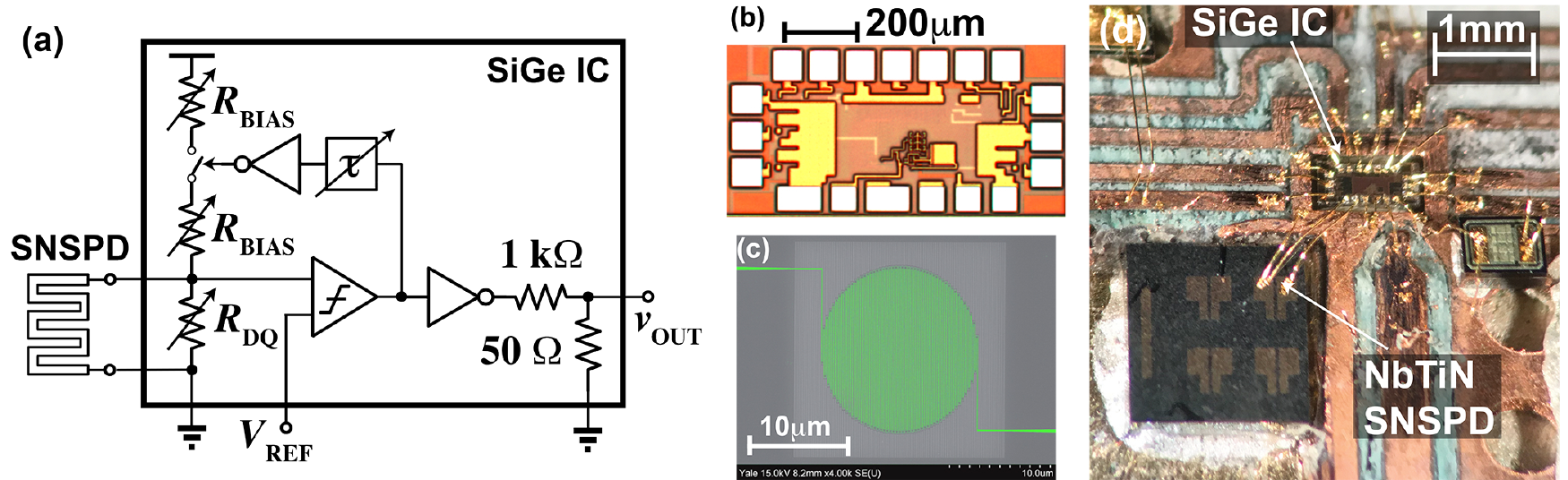}
\caption{Device developed for demonstration of active quenching. (a) Simplified block diagram of SiGe IC. (b) Die photograph of fabricated SiGe IC. {The chip dimensions are $1\,\mathrm{mm}\times{}0.46$\,mm.} (c) False color SEM photograph of example NbTiN detector with active area diameter of \hrd{15\,$\mu$m}. The nanowire width is 50\,nm and the fill factor is 33\%. (d) Photograph of hybrid assembly. The interconnection between the terminals of the SNSPD were directly bonded to the integrated circuit. \hrd{Scale bars in (b)--(d) are approximate.}}
\label{newSCH}
\end{figure}

A proof-of-concept bias and control circuit has been implemented using a commercial \hrd{silicon germanium (SiGe)} BiCMOS  integrated circuit process\cite{orner20030}. A block diagram of the chip connected to an SNSPD appears in Fig.~\ref{newSCH}(a). 
The detector is biased from a resistive current source, which can be disabled using a CMOS switch. The bias current can be controlled using a pair of digitally programmable resistors, each covering the range of 1.2--50\,k$\Omega$. 
The state of the detector is sensed using a comparator whose threshold is set by an off-chip reference voltage. The enable port of the current source is controlled through a delayed version of the comparator output. The delay block was implemented to be asymmetric, such that it mainly affects the falling edge (\hrd{which enables} the bias current \hrd{after quenching has occurred}). The delay associated with the falling edge is programmable over the range of \hrd{0--20}\,ns, \hrd{not including the propagation delay associated with the other components in the feedback loop}. An additional programmable resistor ($R_\text{DQ}$) covering the range of 2--25\,k$\Omega$ was employed to reduce ringing during the quenching and re-biasing phases of operation. For the purpose of this proof-of-concept demonstration, this resistor was not open-circuited during the detection phase.  Finally, a digital output is provided via an inverter that is terminated in a \hrd{41:1} voltage attenuator. This \hrd{attenuator} serves to reduce the power required to drive a 50\,$\Omega$ line. A die micrograph of the fabricated IC appears in Fig.~\ref{newSCH}(b).

A module was designed to allow for evaluation of the active quenching scheme using NbTiN detectors \hrd{with a diameters of 5\,$\mu$m and 10\,$\mu$m}, corresponding to kinetic inductances of 250\,nH and 1\,$\mu$H, respectively. These devices were fabricated on top of a 330\,nm Si$_3$N$_4$ film, which was deposited on an oxidized silicon wafer. The nanowire width was approximately 50\,nm and the NbTiN film thickness, as determined through TEM imaging, was approximately 6.5\,nm. A SEM photograph of a representative device appears in Fig.~\ref{newSCH}(c). Details regarding the nanowire fabrication have been reported previously~\cite{Risheng}. \hrd{{All SNSPDs employed here were taken from the same wafer, which was found to have near uniform performance.}}

From simulation, we estimate the input capacitance of the active quenching IC to be approximately 70\,fF. To avoid the introduction of significant additional stray capacitances, we mounted the detectors in close proximity to the  \hrd{BiCMOS integrated circuit} and made direct bondwire connections between the two chips (see Fig.~\ref{newSCH}(d)). 
Light was coupled to the detectors using flood illumination via SMF-28 fibers that were terminated in open-ended FC/PC connectors. The FC/PC connectors were seated in mating sleeves that were coarsely aligned to the detectors under a microscope. The estimated working distance between the tip the FC/PC connector and the surface of the detector chip was 15\,mm, corresponding to a spot with a diameter on the order of 2\,mm. \hrd{As such, we estimate  geometric coupling losses of 58\,dB and 52\,dB when using 250\,nH (5\,$\mu$m diameter) and 1\,$\mu$H (10\,$\mu$m diameter) detectors, respectively. This large coupling loss was accepted for this proof-of-concept work to ease the challenge of aligning a fiber to the SNSPD.} 

\begin{figure}
\includegraphics[width=\columnwidth]{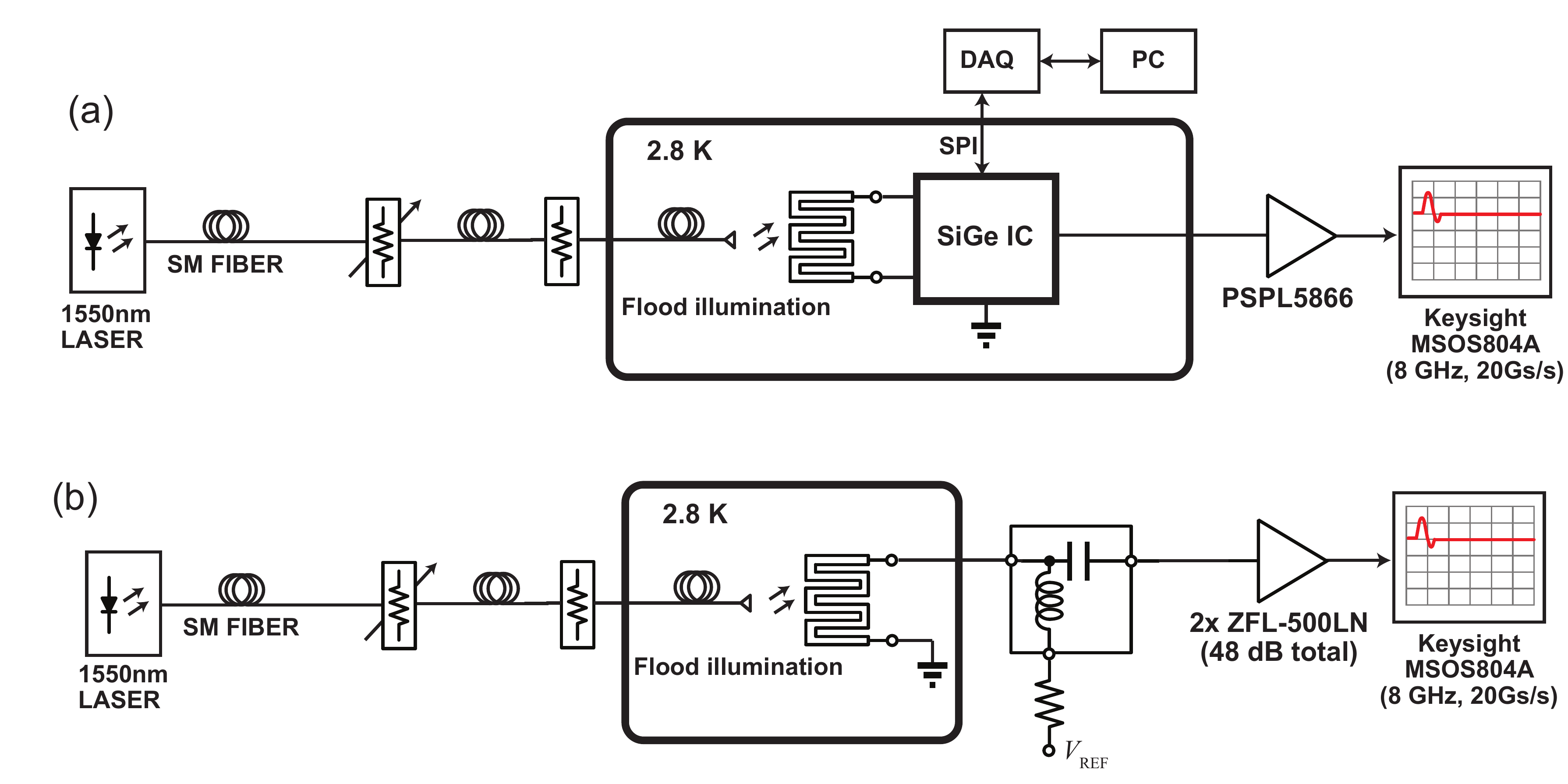}
\caption{Test setup for characterization of the detectors in the (a) active and (b) passive quenching configurations.}\label{setup}
\end{figure}

\hrd{Testing was carried out in a commercial closed-cycle cryostat. The detectors were illuminated} at 1550\,nm using both CW (Ando AQ8204\hrd{, {10}\, mW}) and femtosecond lasers (Calmar FPL-02CFF: \hrd{2\,mW average power, <500\,fs duration, and 30\,MHz repetition rate), attenuated to single photon levels\footnote{\hrd{As an example, to achieve count rates of 200 kcps using the CW and fs laser, we employed 50.5 and 44\,dB of external attenuation, respectively, when the 1\,$\mu$H detector was biased at 95\% of its switching current. Including the estimated 52\,dB of coupling losses, we find a photon flux of about $4\times10^6$ photons/second was incident upon the detector in each case. In the case of the CW laser, we can estimate the probability of a two-photon event by assuming a Poissonian process and a conservative detection window of 1\,ns, we find that there is a 0.4\% probability of a two photon event. Similarly, applying coherent statistics for the pulsed case, we find a 0.8\% chance of more than one photon being incident upon the detector during any given pulse. As these probabilities are both considerably smaller than the estimated system detection efficiency of just over 4\%, we believe that these power levels are well within the single photon regime.}}}. An additional set of baseline measurements were carried out in which the electrical connections between each detector and the active quenching circuit were broken and the detectors were directly wirebonded to the output transmission lines \hrd{of the printed circuit board}. The performance of each device was then measured using a typical 50\,$\Omega$ readout chain. \hrd{This procedure permitted a direct comparison of active and passive quenching with the same detector employed for both sets of measurements}. Detailed block diagrams of the test setups employed for characterization of the detectors in the active and passive quenching configurations appear in Figs.~\ref{setup}(a) and \ref{setup}(b), respectively. 

All measurements were taken at a physical temperature of 2.8\,K, as measured using a silicon diode temperature sensor that was directly mounted on the lid of the modules. Moreover, \hrd{unless otherwise stated}, the active quenching circuit was configured with {$R_\text{DQ}=3.8\,\mathrm{k}\Omega$, $R_\text{BIAS}=10\,\mathrm{k}\Omega$,} and with a reset delay in the range of {4--10\,ns}. The quiescent SNSPD bias current was adjusted using a laboratory voltage source and the active quenching circuit was biased at a power consumption of approximately 100\,$\mu$W, \hrd{excluding} the dissipation of the output driver.

The latching ($I_\text{LATCH}$) and hot-spot currents ($I_\text{SS}$) were measured for \hrd{two of each detector value.} 
\hrd{The latching currents were determined with all fiber ports of the cryostat terminated with metal caps and correspond to the minimum bias currents that caused the detectors to latch due to a dark count.} 
When biased through the active quenching circuit, we measured \hrd{$I_\text{LATCH}=13.1$ and 12.9\,$\mu$A for the 250\,nH detectors and 11.5\,$\mu$A, for both of the 1\,$\mu$H detectors}. On the other hand, when using the passive quenching configuration, we found that the latching currents of the same \hrd{pair of 250\,nH detectors dropped to 12.3\,$\mu$A and {12.1}\,$\mu$A, respectively}. \hrd{Similarly, we found that the latching currents for the pair of 1\,$\mu$H detectors dropped to 10.2 and 11.3\,$\mu$A, respectively.} Interestingly, we found the values of $I_\text{SS}$ to be consistent between the  active and passive quenching configurations, \hrd{with values of 1.86\,$\mu$A and 1.84\,$\mu$A recorded for the 250\,nH detectors and 1.68 and 1.78\,$\mu$A recorded  for the 1\,$\mu$H detectors.} This indicates that the change in latching current was not due to a difference in bath temperature\hrd{, but rather the quenching mechanism.} 

\begin{figure}\centering	
\includegraphics[width=\columnwidth]{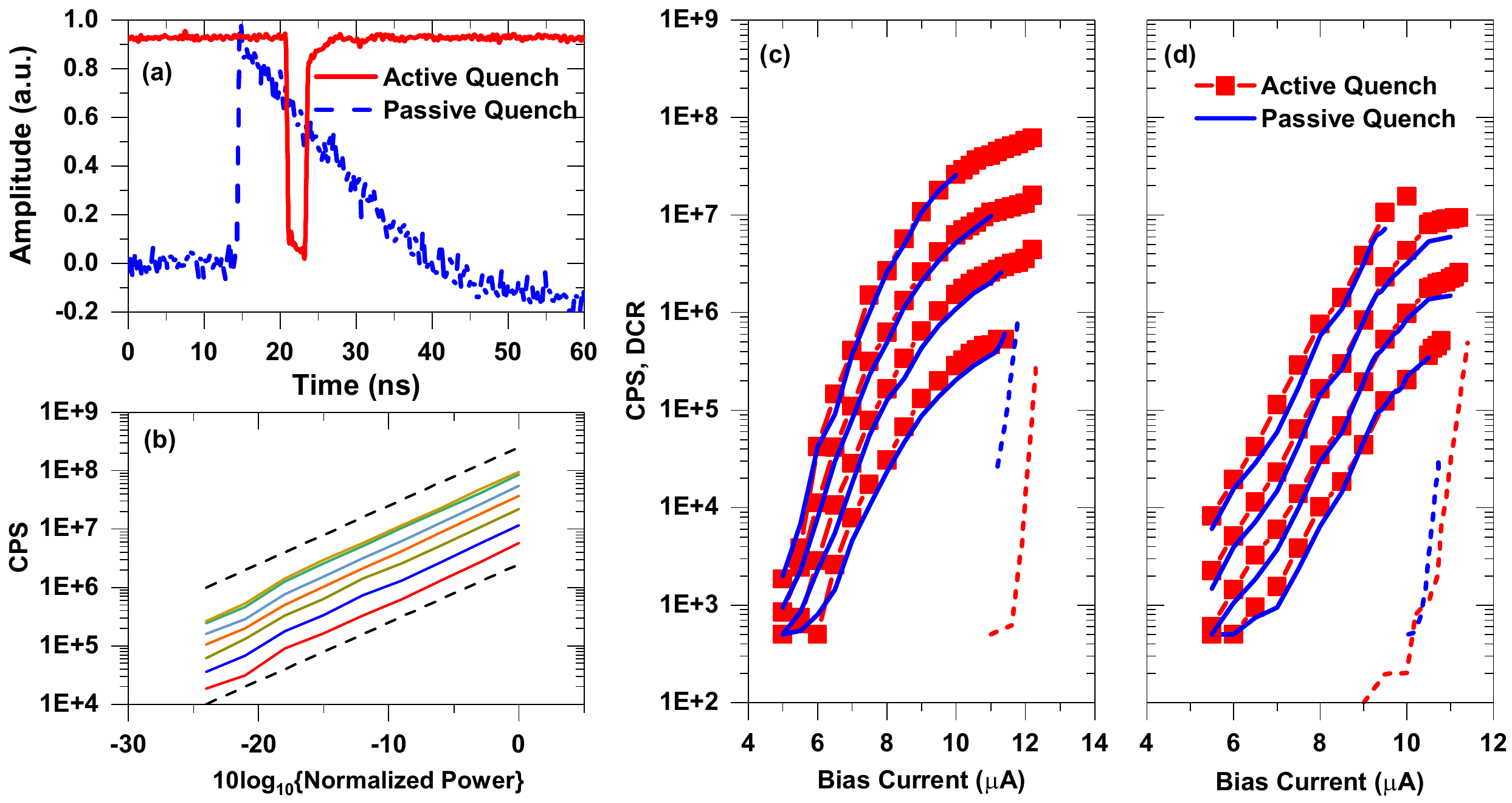}

\caption{\hrd{Results: (a) Time domain waveforms for the 1\,$\mu$H 
detector. (b) Count rate for actively quenched 250\,nH SNSPD at bias currents of 8.0, 8.5, 9.0, 9.5, 10.0, 11.0, and 11.4\,$\mu$A. The count rate was found to increase monotonically with bias current. The dashed lines are guide lines with a slope of one. 
 Count rates for (c) 250\,nH and (d) 1\,$\mu$H 
detectors in the active and passive quenching configurations at four different signal intensities (3\,dB increment). Dark count rates are also shown  as dashed lines for each configuration.}}
\label{EXP}
\end{figure}
Next, we measured the detector performance under CW illumination. Example time domain waveforms measured using each quenching scheme appear in Fig.~\ref{EXP}(a). The output of the active quenching circuit is an inverted pulse with a duration of approximately \hrd{{2.5}}\,ns, which is significantly shorter than the pulse duration observed using the passive quenching scheme. \hrd{This duration is determined by the latency around the feedback loop and could be further reduced by moving to a more advanced CMOS technology node or dissipating more power}. From the measured passive quenching waveforms, we estimated recovery time constants  ($\tau_\text{e}=L_\text{K}/R_\text{L}$) of 20\,ns and 5\,ns for 250\,nH and 1\,$\mu$H devices, respectively. These numbers are consistent with expectation.  


The count rates were also measured under CW illumination. Curves were collected as a function of both detector bias and light intensity. Example results appear in Figs.~\ref{EXP}(b)--(d). In the active quenching configuration, we observed a linear relationship  between the intensity and count rates over a wide range of bias currents and count rates. We also found that count rates greater than 95\,Mcps and 12\,Mcps were achievable for 250\,nH and 1\,$\mu$H devices, respectively. In each case, the device eventually latched as the light intensity was increased. We believe that this effect could be avoided by open-circuiting $R_\text{DQ}$ during the detection phase of operation.

Count rates were also measured with each SNSPD configured in the passive quenching configuration (see Figs.~\ref{EXP}(c) and \ref{EXP}(d)). The results were found to be consistent with the active quenching measurements under low-illumination. However, the maximum count rates that could be achieved before the devices latched were found to be 25\,Mcps and \hrd{7}\,Mcps for 250\,nH and 1\,$\mu$H devices, respectively. 
We believe that the active quenching approach was able to reach higher count rates due to the reduced dead time and dc coupling.



\begin{figure}[bt!]
\includegraphics[width=\columnwidth]{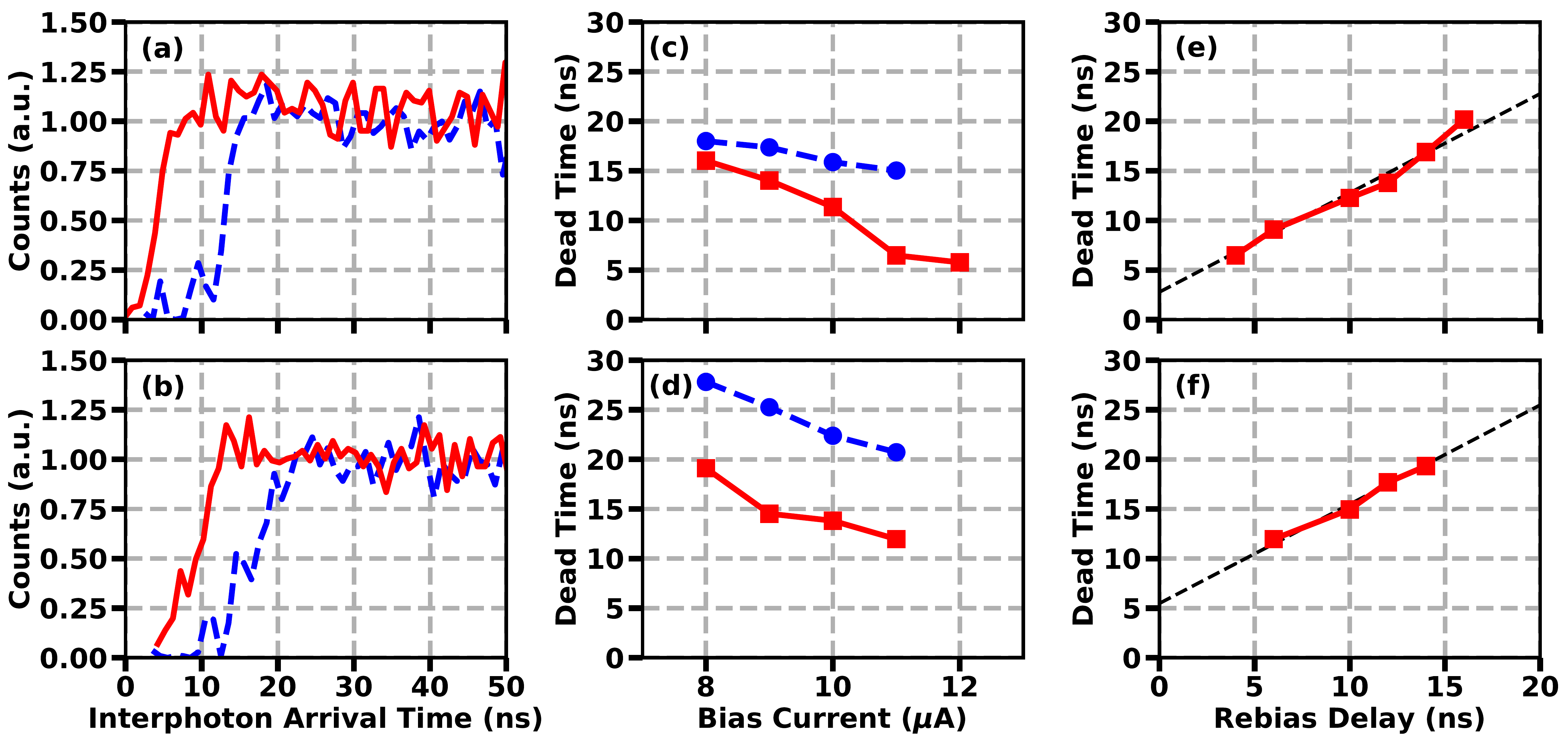}
\caption{\hrd{Interphoton arrival statistics. Example interphoton arrival statistics for (a) 250\,nH and (b) 1\,$\mu$H detectors at a bias current of 11\,$\mu$A. Deadtime as a function of bias current for (c) 250\,nH and (d) 1\,$\mu$H detectors. The solid red and dashed blue lines correspond to data acquired using active and passive quenching, respectively. Curves (a)--(d) were acquired with rebias delays of 4\,ns and 6\,ns for the 250\,nH and 1\,$\mu$H detectors, respectively. Dead time as a function of rebias delay at a bias of 11\,$\mu$A for (e) 250\,nH and (f) 1\,$\mu$H detectors. The solid red and black dashed lines correspond to experimental data and the simple model of dead time on delay dependence, as described in the text.}}\label{interphoton}
\end{figure}

\hrd{
The dead time was studied by performing inter-photon arrival experiments. For these measurements, the devices were excited by a 1550\,nm CW source, attenuated to produce average count rates of approximately 200\,kcps.  Interphoton arrival statistics were acquired over a wide range of detector bias currents and in excess of 500,000 statistics were acquired for each measurement. Unless otherwise stated, the rebias delays were set to {4}\,ns and {6}\,ns for  250\,nH and 1\,$\mu$H, respectively. In all cases, we observed exponential inter-photon arrival histograms whose distributions were consistent with expectation given the observed count rates.

Example inter-photon arrival histograms for 250\,nH and 1\,$\mu$H detectors at a bias current of 11\,$\mu$A are shown in Figs.~\ref{interphoton}(a) and \ref{interphoton}(b), respectively. These data have been normalized by the expected value at zero delay ($2N\sinh\left\{R\Delta\tau/2\right\}$, where $N$ is the total number of counts, $\Delta{\tau}$ is the bin size, and $R$ is the count rate.  The active quenching scheme was found to reduce the recovery time significantly.}

\hrd{To further understand the recovery operation, we quantified the dead time, which we define here as the time after a detection at which the count statistics returned to 90\% of the expected value (assuming Poissonian arrival statistics).  The dead time as a function of detector bias current  is shown for 250\,nH and 1\,$\mu$H devices in Fig.~\ref{interphoton}(c) and (d). respectively. In comparison to passive quenching, the active quenching scheme achieved significantly shorter dead times across the full range of bias currents for which we performed measurements.} 

\hrd{We also studied the dependence of the recovery time on the rebias delay (see Figs.~\ref{interphoton}(e) and (f)). It was possible to acquire data for rebias delays as short as 4\,ns and 6\,ns for  250\,nH and 1\,$\mu$H devices, respectively. At shorter re-bias delays, the devices displayed afterpulsing and latching. Also shown in Figs.~\ref{interphoton}(e) and (f) are trendlines that were generated by assuming that the recovery time was equal to the programmed delay plus a constant offset corresponding the average difference between the measured recovery time and the programmed delay. Referring to Figs.~5(e) and (f), the measured dead times are in fact consistent with the trend lines, validating this simple model. Extrapolation to zero rebias delay indicates recovery time limits of 2.8\,ns and 5.5\,ns for  250\,nH and 1\,$\mu$H detectors, respectively.} 

\hrd{As described earlier, after-pulsing is a potential issue with active quenching, since the dynamic  current can ring and exceed the critical current if the nanowire is rebiased too quickly. To check for after-pulsing, we repeated the inter-photon arrival experiment over the range of bias currents reported in Fig.~\ref{interphoton}, using the femtosecond laser as the excitation source. For these measurements, the optical signal was attenuated to provide a count rate on the order of 200\,kHz when each device was biased at 11\,$\mu$A}. \hrd{In each case, we found that all significant peaks in the interphoton arrival histogram were constrained to time bins that were an integer multiple of the laser period, implying that no significant afterpulsing occurred.}

Next, we measured the dark count rates as a function of bias current. For these measurements, all fiber ports were blocked using metal caps. The results appear in Figs.~\ref{EXP}(c) and \ref{EXP}(d). In each case, we observed significantly lower dark count rates when using active quenching as opposed to passive quenching. Interestingly, this was due to the fact that we observed after-pulsing for each dark count when passive quenching was employed (see Appendix B). No such after-pulsing was observed for photon-induced counts. 
While further work is required to understand this result, one possible explanation could be that the active quenching scheme provides a more effective reset of the detector by forcing the detector bias current to drop below $I_\text{SS}$ for a controlled period of time. 

\begin{figure}[bt!]
\centering
\includegraphics[width=\columnwidth]{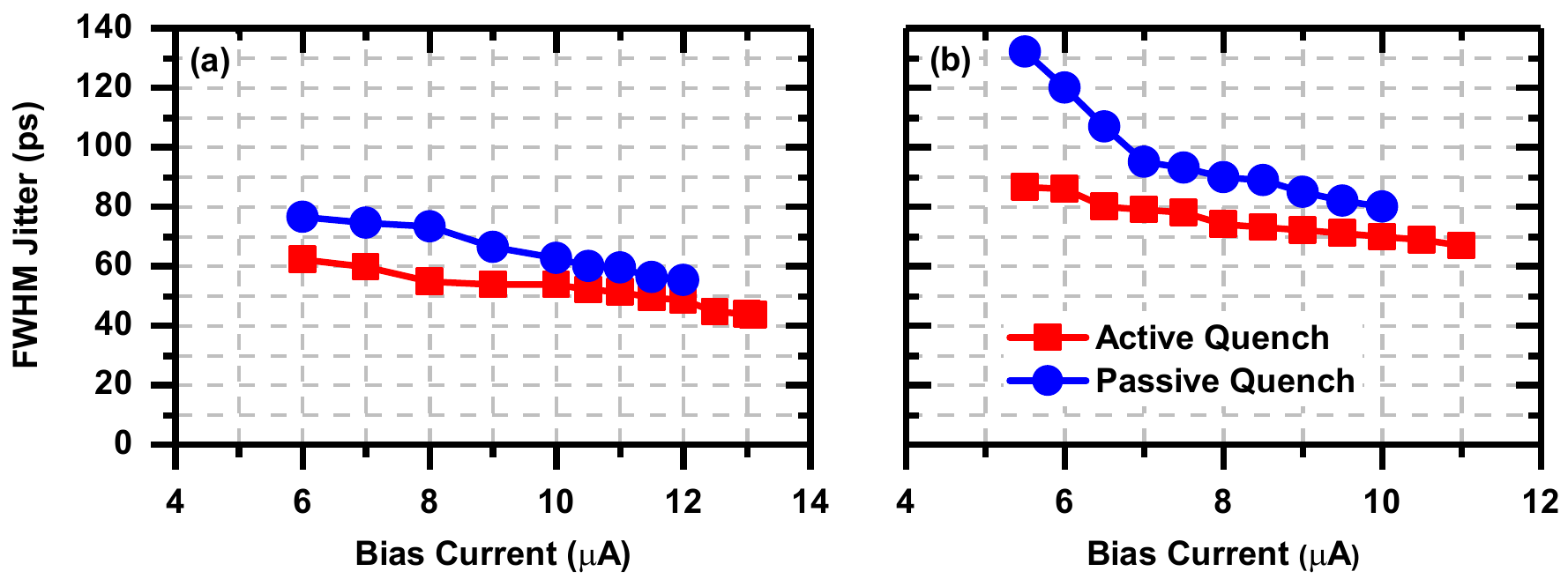}
\caption{Timing jitter as a function of bias current for the (a) 250\,nH and (b) 1\,$\mu$H devices. Improved timing jitter was observed when active quenching was employed.}\label{jitter}
\end{figure}
Finally, we measured the timing jitter of the SNSPDs in each configuration using the 1550\,nm femtosecond laser. The illumination employed for these measurements was set such that count rates of approximately 200\,kHz were achieved when the detectors were biased near their latching currents. \hrd{This count rate is $150\times$ smaller than the repetition rate of the laser and was chosen to ensure that the illumination was in the single photon regime.} The results appear in \hrd{Figs.~\ref{jitter}(a) and \ref{jitter}(b)}. At all bias points, we found that the proposed approach achieved superior jitter performance in comparison to the standard approach. \hrd{As explained in Appendix~C, the difference in measured jitter is not explained by room temperature amplifier noise}.  As the comparator used in prototype circuit was not designed with noise performance in mind, there is likely room for further improvement in the jitter by \hrd{further engineering of this block}.
\section{Conclusion}
Active quenching appears to be a promising approach towards improving the performance of SNSPDs. \hrd{In this article,} we have developed a basic understanding of the performance expected when using active quenching for the bias and readout of an SNSPD and \hrd{have} demonstrated a compact and low-power actively quenched SNSPD circuit. Moreover, we have found experimentally that this bias and readout technique is able to simultaneously improve the count rates, dark count rates, and timing jitter of SNSPDs. 

\hrd{While the results so far are encouraging, further work is required to realize the full potential of actively quenched SNSPDs. For instance, while the jitter of the actively quenched SNSPDs was found to be  lower than that which was achieved when the same detectors were passively quenched, the observed values of jitter were still considerably higher than the current state of the art, which was achieved using passive quenching and cryogenic low noise amplifiers~\cite{JPL}. Further work is required in order to explore the fundamental limits to the jitter of actively quenched SNSPDs. This could, for instance, include the implementation of an active quenching circuit with an ancillary linear buffer amplifier to permit direct monitoring of the voltage across the detector.}

\hrd{Another area in which there is room for improvement is in reducing parasitic capacitances, which slow the achievable slew rate of an actively quenched SNSPD. For this preliminary work, we used relatively large bondpads on both the superconductive and BiCMOS ICs and connected the two using wire bonds, resulting in a total parasitic capacitance that is on the order of the effective normal domain capacitance. A much more optimum approach would be to either fabricate the SNSPD directly on top of the silicon IC, removing the need for large bondpads, or to use membrane SNSPDs~\cite{Faraz} to allow for co-integration.}

\hrd{Finally, for this proof-of-concept work, we have made no attempt to couple light efficiently to the SNSPD. However, if active quenching is to be used in a practical application, it is essential that light be efficiently coupled to the detector. While it is important that the control electronics and the detector are in close proximity to minimize parasitic capacitance, we do not believe this precludes efficient fiber coupling to the detector. Specifically, we believe that it should be feasible to employ either the self-aligned technique described in~\cite{SaeWooSelfAlign} or waveguide coupling~\cite{Hong}. We note that, to make the former approach work assuming separate superconductor and semiconductor chips, it may be necessary to space the detector approximately 1.5\,mm from the BiCMOS IC. However, it should still be feasible to make an interconnect between the two devices using superconductive high impedance (inductive) transmission lines. 

In summary, we have shown via theoretical analysis and proof-of-principle demonstration that active quenching of SNSPDs provides performance advantages in terms of dead time and jitter. Future work in this area should focus on improved performance, with potential research directions including lower noise integrated electronics, tighter semiconductor/superconductor integration, and optimized fiber coupling.}

\section*{Appendix A: Derivation and validation of Equations (1) and (2)}
Here, we derive equations for the ratio of rising-edge slew rate and peak voltage for the active quenching topology with respect to a passive quenching approach. \hrd{In doing so, we neglect the current-dependence of the kinetic inductance~\cite{clem2012kinetic} as well as distributed effects~\cite{santavicca2016microwave}}. We first derive expressions for the slew rate, rise time, and peak voltage for the active quenching architecture. Next, we repeat this exercise for a passively quenched SNSPD and take ratios to arrive at  Equations~(1) and (2) from the main text. Finally, the expressions are validated through circuit simulations.
\subsection*{Rising-edge dynamics of an actively quenched SNSPD}
The rising-edge dynamics of an actively quenched SNSPD can be studied by considering the case of a capacitively terminated SNSPD, biased with a constant current source. As described in the main text, the normal domain dynamics of the detector embedded in this \hrd{circuit} can be modelled using a non-linear capacitance, $C_\text{eff}\left(i_\text{D}\right)$, where $i_\text{D}$ is the instantaneous current flowing through the normal domain (see the main text for the functional form of $C_\text{eff}\left(i_\text{D}\right)$). 

We begin by estimating the average output slew rate of the voltage across the detector. We note that the currents through the normal domain at $t=0$ and $t=t_\text{D}$ are $I_\text{B0}$ and $I_\text{SLEW}$, respectively.  Using this information, we take a linear approximation for the non-linear capacitance such that its average value between $t=0$ and $t=t_\text{D}$ is $\left<C_\text{eff}\right>\approx{\left(C_\text{eff}\left\{I_\text{B0}\right\}+C_\text{eff}\left\{I_\text{SLEW}\right\}\right)}/{2}$.  The average output slew rate is thus approximated as 
\begin{equation}\left<SR_\text{AQ}\right>\approx{I_\text{B0}/\left(\left<C_\text{eff}\right>+C_\text{L}\right)}.\label{SR}\end{equation} 
Next, we write the output voltage at the point in time at which the inductor has fully discharged ($t=t_\text{D}$) as 
\begin{equation}
v_\text{PK,AQ}\approx{}R_\text{NPK,AQ}I_\text{SLEW},
\end{equation}
where $R_\text{NPK,AQ}$ is the normal domain resistance at $t=t_\text{D}$. Since we also know that $v_\text{PK,AQ}\approx{}\left<SR_\text{AQ}\right>t_\text{rise,AQ}$, we can write an expression for the peak normal domain resistance in terms of the rise time
\begin{equation}
R_\text{NPK,AQ}\approx\frac{\left<SR_\text{AQ}\right>t_\text{rise,AQ}}{I_\text{SLEW}}.
\end{equation}
Now, assuming that the dominant impedances in the circuit during the discharge period are the kinetic inductance and the normal domain resistance, the discharge time can be estimated as $t_\text{rise,AQ}\approx{}2.2L_\text{K}/\left<R_\text{N,AQ}\right>$. Finally, taking $\left<R_\text{N,AQ}\right>\approx{}R_\text{NPK,AQ}/2$, we arrive at expressions for the discharge time and peak output voltage,
\begin{equation}
t_\text{rise,AQ}\approx{}\sqrt{2.2\frac{I_\text{SLEW}}{I_\text{B0}}L_\text{K}\left(\left<C_\text{eff}\right>+C_\text{L}\right)}
\end{equation}
and
\begin{equation}
v_\text{PK,AQ}\approx{}\sqrt{\frac{2.2L_\text{K}I_\text{B0}I_\text{SLEW}}{\left<C_\text{eff}\right>+C_\text{L}}}.
\end{equation}
\subsection*{Rising-edge dynamics of a resistively shunted SNSPD}
Next, we approximate the performance characteristics of a resistively shunted SNSPD so that we can make a comparison to the proposed technique. To begin, we note that the current through a resistively shunted SNSPD drops from $I_\text{B0}$ to approximately zero during the time period where the output is transitioning. During this time, a normal domain forms and a voltage develops across this normal domain. Rather than approximating the resistance and current with linear functions, as we did for the case of a capacitively shunted SNSPD, we use exponential functions since the dynamics in this case are more strongly dominated by the kinetic inductance. Specifically, we model the normal domain current and resistance as
\begin{equation}
I_\text{D,PQ}\approx{}I_\text{B0}\exp\left\{\frac{-2.2t}{t_\text{rise,PQ}}\right\}
\label{pcurrent}
\end{equation}
and
\begin{equation}
R_\text{N,PQ}\approx{}R_\text{NPK,PQ}\left(1-\exp\left\{\frac{-2.2t}{t_\text{rise,PQ}}\right\}\right),
\label{pres}
\end{equation}
\hrd{where $I_\text{D,PQ}$ is the dynamic nanowire current, $t_\text{rise,PQ}$ is the rise time of the passively quenched detector; and $R_\text{N,PQ}$ and $R_\text{NPK,PQ}$ are the dynamic and peak normal domain resistances, respectively}. Taking the derivative of the product of Equations.~(\ref{pcurrent}) and (\ref{pres}) with respect to time, we arrive at an expression for the rate at which the voltage across the normal domain grows.
\begin{equation}
SR_\text{N,PQ}\left(t\right)\approx\frac{2.2R_\text{NPK,PQ}I_\text{B0}}{t_\text{rise,PQ}}\left(2\exp\left\{\frac{-4.4t}{t_\text{rise,PQ}}\right\}-\exp\left\{\frac{-2.2t}{t_\text{rise,PQ}}\right\}\right).
\label{SRN}
\end{equation}
At $t=0$, we know that the voltage across the normal domain will slew as $SR_\text{N,PQ}\left(0\right)\approx{}I_\text{B0}/C_\text{eff}\left\{I_\text{B0}\right\}$. Thus, 
\begin{equation}
t_\text{rise,PQ}\approx{}2.2R_\text{NPK,PQ}C_\text{eff}\left\{I_\text{B0}\right\}.
\end{equation}
We can also approximate the rise time in terms of the kinetic inductance and average normal domain resistance as $t_\text{rise,PQ}\approx{}2.2L_\text{K}/\left<R_\text{N,PQ}\left(t\right)\right>$, where 
\begin{equation}
\left<R_\text{N,PQ}\left(t\right)\right>\approx\frac{R_\text{NPK,PQ}}{t_\text{rise,PQ}}\int_0^{t_\text{rise,PQ}}\left(1-\exp\left\{-2.2t/t_\text{rise,PQ}\right\}\right)\mathrm{dt}.
\end{equation}
Evaluating this integral, we find $\left<R_\text{N,PQ}\right>\approx{}3R_\text{NPK,PQ}/5$. Thus, the peak normal domain resistance, the rise time, the peak output voltage voltage, and the average slew rate can be approximated as 
\begin{equation}
R_\text{NPK,PQ}\approx\sqrt{\frac{5}{3}\frac{L_\text{K}}{C_\text{eff}\left\{I_\text{B0}\right\}}},
\end{equation}
\begin{equation}
t_\text{rise,PQ}\approx{}2\sqrt{2L_\text{K}C_\text{eff}\left\{I_\text{B0}\right\}},
\end{equation}
\begin{equation}
v_\text{PK,PQ}\approx{}I_\text{B0}R_\text{L},
\end{equation}
and
\begin{equation}
\left<SR_\text{PQ}\right>\approx\frac{I_\text{B0}R_\text{L}}{2\sqrt{2L_\text{K}C_\text{eff}\left\{I_\text{B0}\right\}}}.
\end{equation}
\subsection*{Comparison of rising-edge dynamics for active and passive quenching}
Finally, we take the ratios of the active quenching metrics to those of the passive quenching approach in order to quantify the improvement in rise time characteristics.
\begin{equation}
\frac{\left<SR_\text{AQ}\right>}{\left<SR_\text{PQ}\right>}\approx{}\frac{2\sqrt{2L_\text{K}C_\text{eff}\left\{I_\text{B0}\right\}}}{R_\text{L}\left(\left<C_\text{eff}\right>+C_\text{L}\right)},
\label{eq1}
\end{equation}
\begin{equation}
\frac{t_\text{rise,AQ}}{t_\text{rise,PQ}}\approx\frac{1}{2}\sqrt{1.1\frac{I_\text{SLEW}}{I_\text{B0}}\frac{\left<C_\text{eff}\right>+C_\text{L}}{C_\text{eff}\left\{I_\text{B0}\right\}}},
\end{equation}
and
\begin{equation}
\frac{v_\text{pk,AQ}}{v_\text{pk,PQ}}\approx{}\sqrt{2.2\frac{I_\text{SLEW}}{I_\text{BO}}\frac{L_\text{K}/R_\text{L}}{R_\text{L}\left(\left<C_\text{eff}\right>+C_\text{L}\right)}}
\label{eq2}
\end{equation}

\begin{table}[bt!]
\centering
\textbf{\caption{Parameters used for verification of equations (1) and (2), (From \cite{KERMANFB,duan2010sub})}.}
\begin{tabular}{llllllllll}
\hline
&$L_\text{S}$&$h_\text{c}$&$\kappa$&$c$&$\rho_\text{n}$&$T_\text{C}$&$T_\text{SUB}$&$w$&$t$\\
\hline
Units&pH/$\Box$&W$\cdot$m$^{-2}\cdot$K$^{-1}$&\,W$\cdot$m$^{-1}\cdot$K$^{-1}$&J$\cdot$m$^{-3}\cdot$K$^{-1}$&$\Omega\cdot$m&K&K&nm&nm\\
\hline
Value&80&50,000&0.108&4,400&3$\times$10$^{-6}$&10.5&2.5&80&4\\
\hline
\label{Table1}
\end{tabular}
\end{table}

\subsection*{Numerical validation of Equations (\ref{SRcomp}) and (\ref{voltcomp})}
A series of simulations were conducted to validate and interpret Equations ~(\ref{SRcomp}) and (\ref{voltcomp}). These simulations were carried out in a Spice environment using the model topology reported in {\cite{SUSTKB}. The topologies considered are those from Fig.~2 of the main text, with $R_\text{DQ}$ open circuited for the active quenching case. A list of model parameters used for the SNSPD appears in Table~\ref{Table1}.
\begin{figure}
\includegraphics[width=1\columnwidth]{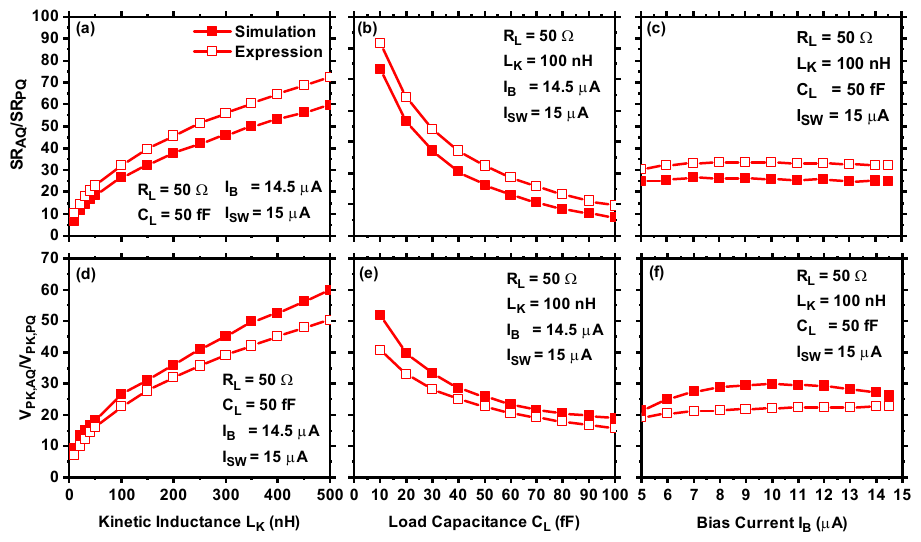}
\caption{\hrd{Verification of equations (\ref{SRcomp}) and (\ref{voltcomp}). Ratio of slew rate for active quenching to that of passive quenching as a function of (a) kinetic inductance, (b) load capacitance, and (c) bias current. Ratio of peak voltage achieved with active quenching to that of passive quenching as a function of (d) kinetic inductance, (e) load capacitance, and (f) bias current. The simulations were carried out using  the model described in~\cite{SUSTKB}. }}
\label{Fig2}
\end{figure}

Simulations were carried out for both the actively quenched and resistively shunted SNSPDs over a \hrd{wide} range of kinetic inductances, load capacitances, and bias currents.  The ratio of slew rate and rise time for the actively quenched SNSPD to that of the resistively shunted device was calculated for each case and the results are shown in Figs.~\ref{Fig2}(a)--\ref{Fig2}(f). While the simulations predict the active quenching architecture will achieve a slightly lower improvement in slew rate and a slightly higher improvement in peak voltage in comparison to the that expected from Equations.~(1) and (2) of the main text, the trends are consistent with expectation. 

Of particular importance is the dependence of the relative improvements of the load capacitance for the actively quenched detector. \hrd{As the load capacitance increases, the improvement which can be achieved using the active quenching architecture decreases.} Thus, it is critical to minimize this capacitance. Since a high impedance comparator has a capacitive input, there is a practical limit to this value\hrd{;  we assume this minima }to be on the order of 30\,fF, which is slightly higher than a typical bondpad capacitance. However, even with a capacitance of this size, the improvement in performance is still significant. 
\section*{Appendix B: Dark count after-pulsing}
The dark count measurements displayed after pulsing when the passive quenching configuration was employed. Such after-pulsing was not observed for regular counts or when using the active quenching configuration. An example transient waveform demonstrating this phenomenon appears in Fig.~\ref{BURST}(a).  Here, a single dark count produces a string of pulses over the course of approximately 0.2\,$\mu$s. This waveform was obtained using a 250-nH detector. However, similar behavior was also observed for the 1-$\mu$H device. As a result of the dark count induced after-pulsing, the recorded dark count rates for the passively quenched devices were significantly greater than those of the active devices. 

\begin{figure}
\includegraphics[width=\columnwidth]{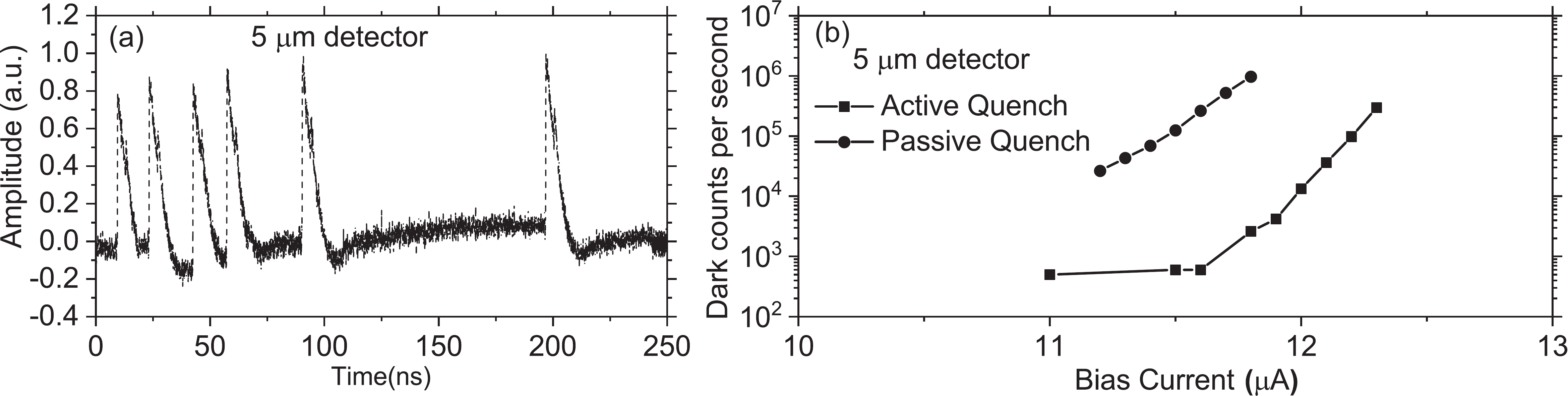}
\caption{~Dark count after-pulsing observed for the passively quenched devices. (a) Time domain waveform for an example dark count and (b) comparison of dark counts for the same detector in the passive and active quenching configurations.}
\label{BURST}
\end{figure}

\section*{Appendix C: Jitter contribution due to room temperature amplifiers}
\label{JitterAPP}
Here, we estimate the jitter contribution from the room temperature amplifiers employed in the passive quenching configuration in order to rule this out as the reason for the improvement in jitter associated with active quenching. The RMS jitter due to these amplifiers can be estimated as the ratio of the RMS noise voltage to the rising-edge slew-rate, both measured at the output of the amplifier. The measured RMS noise voltages and slew rates for a 250\,nH(1\,$\mu$H) detector were found to be 5.3(4.7)\,mV and 580(510)\,MV/s, respectively. Thus, the jitter due to amplifier noise was estimated to be 9\,ps RMS, corresponding to 21\,ps FWHM. Considering the fact that the amplifier jitter contribution should be independent from other sources of jitter, we can de-embed the effect of this jitter: $T_\text{J,int}=\sqrt{T_\text{J,meas}^2-T_\text{J,amp}^2},$ where $T_\text{J,int}$ and $T_\text{J,meas}$ are the intrinsic and measured jitter, and $T_\text{J,amp}$ is the amplifier jitter contribution. Applying this correction to the data shown in Fig.~\ref{EXP}, we find that the amplifier jitter contribution is at most 4\,ps and that the jitter performance of the actively quenched detector is still superior in all cases.  
\section*{Acknowledgement}
This work was supported by the Office of Naval Research (ONR) (\#N00014-15-1-2417), the Defense Advanced Research Project Agency (DARPA) (\#W911NF-16-2-0151), the National Science Foundation (\#CCCS-1351744), and Google LLC.
\\



\bibliographystyle{IEEEtran}
\bibliography{ACTIVE_RESET_arxiv}

\end{document}